\documentclass[prd,a4paper,preprint,preprintnumbers,nofootinbib]{revtex4-2}

\usepackage[a4paper, hdivide={1.91cm,,1.165cm}, vdivide={1.83cm,,3.0cm}]{geometry}

\usepackage{amsmath}
\usepackage{graphicx}
\usepackage{xspace}
\usepackage{color}
\usepackage{units}
\usepackage{slashed}
\usepackage{gensymb}
\usepackage{booktabs}
\usepackage{multirow}
\usepackage{xcolor}
\usepackage{setspace}
\usepackage{booktabs}
\usepackage{hyperref}
\hypersetup{
colorlinks=true,
urlcolor=blue,
citecolor=blue}

\usepackage{xspace}
\usepackage{xcolor}
\usepackage{setspace}
\usepackage{mathtools}

\usepackage{physics}
\usepackage{nicefrac}

\usepackage{bbold}

\usepackage{accents}

\newcommand{\jp}{J^{P}}

\newcommand{\tablexi}[2]{$\Xi^{#1}_{#2}$}
\newcommand{\tableom}[2]{$\Omega^{#1}_{#2}$}

 \allowdisplaybreaks
\newcommand{\prop}[2]{S^{#1}_#2(x)} 

\newcommand{\tilprop}[2]{\widetilde{S}^{#1}_#2(x)}

\newcommand{\gf}{\gamma_5}

\newcommand{\inp}[1]{\left(#1\right)}

\newcommand{\inb}[1]{\left[#1\right]}

\begin{document}


\title{Properties of doubly heavy baryons in QCD}

\author{T.~M.~Aliev}
\email{taliev@metu.edu.tr}
\affiliation{Department of Physics, Middle East Technical University, Ankara, 06800, Turkey}

\author{S.~Bilmis}
\affiliation{Department of Physics, Middle East Technical University, Ankara, 06800, Turkey}
\affiliation{TUBITAK ULAKBIM, Ankara, 06510, Turkey}

\begin{abstract}
The study of the properties of doubly heavy baryons represents a promising area in particle physics. It can provide us with information about Cabibbo--Kobayashi--Maskawa (CKM) matrix elements and the low energy dynamics of QCD. They have a very rich phenomenology. The investigation of weak, electromagnetic, and strong decays has a vital importance for understanding the dynamics of doubly heavy baryons. The main ingredients of such studies are the spectroscopic parameters, the strong coupling constants, and the transition form factors. For calculations of these quantities, non-perturbative methods are needed. One of these methods is the QCD sum rules. In the present work, we review our studies on the properties of the doubly heavy baryons within the sum rules method, focusing mainly on the mass and strong coupling constants of the doubly heavy baryons. In addition, the radiative decays of doubly heavy baryons are estimated using the vector dominance model. We also make few remarks on the semileptonic decays of the doubly heavy baryons.
\end{abstract}
\maketitle

\newpage

\section{Introduction\label{intro}}
The constituent quark model has been a powerful tool for studying the hadron spectroscopy~\cite{GellMann:1964nj}. The model predicts the existence of hadrons containing single, double, and triple heavy quarks. Even though many baryons with single heavy quarks (even the excited states) like $\Omega_b$, $\Xi_b$, $\Sigma_b$, $\Lambda_b$, $\Omega_c$, $\Xi_c$, and $\Lambda_c$ have been discovered by several flavor and accelerator experiments like CDF, CLEO, BABAR, BELLE and LHCb~\cite{ParticleDataGroup:2020ssz}, search for the baryons involving two heavy quarks ended with no observation ~\cite{BaBar:2006bab,Belle:2013htj,LHCb:2013hvt} for a long time. This has been a long-standing puzzle for the quark model. In 2002, SELEX collaboration made a breakthrough with the discovery of $\Xi_{ccd}^+ (3520)$ in $p D^+ K^-$ decay channel~\cite{SELEX:2002wqn} and the confirmation was also done by the same experiment in 2005~\cite{SELEX:2004lln}. Moreover, in 2017, LHCb collaboration announced the discovery of $\Xi_{ccu}^{++} (3621)$ via $\Xi_c^{++} \rightarrow \Lambda_c^+ K^- \pi^+ \pi^+$ decay~\cite{LHCb:2017iph}. This state was also confirmed by LHCb collaboration in 2018 in the decay $ \Xi_{cc}^{++} \rightarrow \Xi_{c}^{+} \pi^{+} $ with $5.9 \sigma$ statistical significance~\cite{LHCb:2018pcs}. Experimental studies have been conducted to unveil the characteristic properties of this new baryon~\cite{LHCb:2018zpl,LHCb:2019epo,LHCb:2019ybf} as well as searching new doubly heavy baryons~\cite{LHCb:2021xba}. 

Many theoretical studies inspired by these discoveries also have been carried out to determine the characteristic properties like the masses, lifetimes, strong coupling constants, and decay widths of doubly heavy baryons by investigating the weak, electromagnetic, and strong decays of these baryons within various models \cite{Alrebdi:2020rev,Rostami:2020euc,Aliev:2020aon,Aliev:2021hqq,Azizi:2020zin,Olamaei:2020bvw,Yu:2017zst,Luchinsky:2020fdf,Gerasimov:2019jwp,Wang:2017mqp,Meng:2017udf,Gutsche:2017hux,Xiao:2017udy,Xiao:2017udy,Lu:2017meb,Xiao:2017dly,Zhao:2018mrg,Xing:2018lre,Jiang:2018oak,Gutsche:2018msz,Dhir:2018twm,Gutsche:2019wgu,Yu:2019yfr,Gutsche:2019iac,Ke:2019lcf,Cheng:2020wmk,Hu:2020mxk,Li:2020qrh,Han:2021gkl,Wang:2017azm,Shi:2017dto,Zhang:2018llc,Ivanov:2020xmw,Rahmani:2020pol,Li:2017pxa,Berezhnoy:2018bde,Guo:2017vcf,Ma:2017nik,Yao:2018zze,Yao:2018ifh,Meng:2018zbl,Shi:2020qde,Qiu:2020omj,Olamaei:2021hjd,Qin:2021dqo,Hu:2017dzi,Li:2018epz,Shi:2019hbf}. Investigating the possible decays of these baryons can give crucial information about the dynamics of these processes. While the transition form factors can be determined by studying the weak and electromagnetic decays, coupling constants can be calculated for analyzing the strong decays.

In the quark model, the baryons are described by quark-diquark picture. Diquarks form a system either with spin-zero or spin-$1$. For this reason, the total spin of the diquark system may be either spin $1/2$ or $3/2$. The lowest-lying doubly heavy baryons with $\jp = 1/2^{+}$ form the SU$(3)$ triplet $\Xi^{++}_{QQ}(QQu)$, $\Xi_{QQ}(QQd)$, and $\Omega_{QQ}(QQs)$. The bottom-charm baryons form two sets of SU$(3)$ triplets ($\Xi_{bc}$,$\Omega_{bc}$) and ($\Xi^{\prime}_{bc}$,$\Omega^{\prime}_{bc}$). The difference between these sets is that while the total spin of the $bc$ system is equal to one for the first set, it is equal to zero for the second case. It is also possible that mixing exists between these two sets. The doubly heavy baryons $\jp = 3/2^{+}$ can decay into the lowest-lying ones either radiatively or with pion emission. In Table~\ref{tab:1}, the baryons with two heavy quarks and with corresponding quantum numbers are represented. 

The weak decays of doubly heavy baryons to the single heavy baryons can be described by spin-$1/2$ to spin-$1/2$, spin-$1/2$ to spin-$3/2$, spin-$3/2$ to spin-$1/2$, and spin-$3/2$ to spin-$3/2$ transitions. 

It is a well-known fact that weak transitions take place via charged or neutral currents. 
\begin{itemize}
    \item The spin-$1/2$ to spin-$1/2$ ($3/2$) transition with charged currents is described by $ c \rightarrow d, s $ or $ b \rightarrow u, c $ processes.
    \item The spin-$1/2 \rightarrow$ spin-$1/2$ ($3/2$) transition via flavour changing neutral current (FCNC) is described by $ c \rightarrow u $ or $ b \rightarrow d,s $ transitions. 
\end{itemize}

All these processes are described by the transition matrix elements between initial and final states. The formation of hadrons from quarks takes place at a large distance which belongs to the nonperturbative domain of QCD. For this reason, for describing these transitions, we need some nonperturbative methods. Among all nonperturbative methods, the sum rules method occupies a special place because it is based on QCD Lagrangian.
\begin{table*}[htb]
  \centering
  \renewcommand{\arraystretch}{1.4}
  \setlength{\tabcolsep}{3.2pt}
  \begin{tabular}{lccc}
    \toprule
        Baryon & $\jp$ & Baryon & $\jp$ \\
    \midrule
        \tablexi{}{QQ}          &   $1/2^{+}$ &   \tableom{}{QQ}        & $1/2^{+}$\\
        \tablexi{*}{QQ}         &   $3/2^{+}$ &   \tableom{*}{QQ}       & $3/2^{+}$\\
        \tablexi{\prime}{bc}    &   $1/2^{+}$ &   \tableom{\prime}{bc}  & $1/2^{+}$\\
        \tablexi{}{bc}          &   $1/2^{+}$ &   \tableom{}{bc}        & $1/2^{+}$\\
        \tablexi{*}{bc}         &   $3/2^{+}$ &   \tableom{*}{bc}       & $3/2^{+}$\\
    \bottomrule
  \end{tabular}
  \caption{Baryons containing two heavy quarks with corresponding quantum numbers are represented. Baryons with prime means that this state is antisymmetric with respect to the flavors, and diquark has spin-zero.}
  \label{tab:1}
\end{table*}
The semileptonic decays of baryons with two heavy quarks take place through the weak current. The weak decays of a doubly heavy baryon to a single heavy baryon transition within the QCD sum rules \cite{Shi:2019hbf}, the light cone version of sum rules \cite{Shi:2019fph, Hu:2019bqj} as well as in the framework of the light-front quark model \cite{Ke:2019lcf, Xing:2018lre, Zhao:2018mrg} are comprehensively studied.

In this work, we review our studies~\cite{Shi:2019fph,Alrebdi:2020rev,Aliev:2020aon,Aliev:2021hqq,Aliev:2012iv,Aliev:2019lvd,Aliev:2012ru} where we estimate the mass and decay constants as well as strong coupling constants of the doubly heavy baryons with light mesons. Moreover, radiative widths are also estimated by using the vector-meson dominance hypothesis.
The paper is organized as follows. In Section~\ref{sec:2}, we briefly discuss the sum rules method. In Section~\ref{sec:3}, we represent the details of the calculations for the mass and decay constants within QCD sum rules. In Section~\ref{sec:4}, we discuss the determination of the strong coupling constants for the  doubly heavy baryons with light pseudo-scalar and vector mesons. Section~\ref{sec:5} is devoted to the our estimation of the radiative decays of doubly heavy baryons by using the vector dominance model and the last section contains our summary and conclusion.

\section{Sum rules method}
\label{sec:2}
The QCD sum rules were first proposed in~\cite{Shifman:1978bx} for mesons and later generalized to baryons in~\cite{Ioffe:1981kw}. This method is powerful for studying the properties of hadrons. Various aspects of this method are discussed in many review papers (see for example \cite{Reinders:1984sr} and \cite{Colangelo:2000dp}).
The main advantage of the sum rule is that it is based on the fundamental QCD Lagrangian. Investigating the bound state problem in QCD, this method starts with short distance relations and then includes the nonperturbative effects of QCD vacuum to probe the large distance.
The correlation function is the essence of the sum rules. It is constructed in terms of the interpolating currents, which carry the same quantum numbers as the hadrons under consideration. The mass and decay constants of the hadrons are usually studied within the so-called two-point correlation function. On the other hand, the strong coupling constants and form factors are investigated within three-point sum rules.

A generic two and three-point correlation function can be written as
\begin{align}
        \Pi(p) &= i\int d^4x e^{ipx} \bra{0}\mathcal{T}\left\{j(x)j^\dagger(0)\right\}\ket{0} \label{201} \\
        \Pi(p,q) &= i^2\int d^4x d^4y e^{ipx+iqy}\bra{0}\mathcal{T}\left\{j(x)j(0)j^\dagger(y)\right\}\ket{0}~.  \label{202}
\end{align}

Moreover, the following correlation function is introduced in so-called light-cone sum rules (LCSR).
\begin{equation}
    \Pi(p) = i\int d^4x e^{ipx}\bra{0}\mathcal{T}\left\{j(x)j^\dagger(0)\right\}\ket{M(q)}~,
    \label{203}
\end{equation}
where $M$ is the corresponding meson, baryon, or photon.

The LCSR was introduced in~\cite{Balitsky:1989ry} in order to solve some drawbacks of three-point sum rules (see \cite{Braun:1997kw}).
The correlation function is calculated in terms of hadrons and quark-gluon degrees of freedom. Matching these two representations, one can obtain the desired sum rules, which allow us to express the properties of hadrons in terms of the quarks and gluons. From the QCD side, the correlation function is calculated with the help of operator product expansion (OPE), which can be represented as a multiplication of the so-called Wilson coefficients and vacuum expectation values of the local operators. In two and three-point sum rules, the OPE is carried out with respect to the dimensions of the local operators.

In light cone sum rules, the OPE is carried out with respect to the twist, which is equal to the difference of the dimension of operators and their spins. In this method, instead of local condensates, which is the main ingredient of the traditional QCD sum rules, they appear as distribution amplitudes (DAs). These DAs contain all the information about the nonperturbative effects.
\section{Calculation of the mass and decay constants}
\label{sec:3}
The mass and decay constants of the baryons can be studied in the framework of the two-point sum rules. For this purpose, we introduce the following two-point sum rules.
\begin{equation}
    \Pi(p) = i \int d^4x e^{ipx} \bra{0} \mathcal{T}\left\{j(x)j^\dagger(0)\right\}\ket{0}
    \label{301}
\end{equation}
To calculate this correlation function, we need to know the form of interpolating current. For the spin-$\nicefrac{1}{2}$ double heavy baryon, we now describe the details of the derivation of the form of the interpolating current. Baryons in quark model is described with the help of diquark-quark picture. A diquark is represented by two heavy quarks. Hence, the interpolating field of the diquark should be similar to the interpolating current for a meson, i.e.
\begin{equation}
    J_{\text{meson}} = \bar{q} \Gamma q~,
    \label{302}
\end{equation}
where $\Gamma$ is the arbitrary Dirac matrix. In the quark model, baryons consist of three quarks. Therefore $\bar{q}$ should be replaced by its charge-conjugation. The charge-conjugated state is related to the $\bar{q}$ as 
\begin{equation}
    q^C = C \bar{q}^T~,
    \label{303}
\end{equation}
where $C$ is the charge conjugation matrix. From the above equation, one gets
\begin{equation}
    \bar{q} = q^TC~.
    \label{304}
\end{equation}
As a result, the interpolating field for a diquark can be written as
\begin{equation}
    J_{diquark} = q^T C q~.
    \label{305}
\end{equation}
The interpolating current for the doubly heavy baryons is constructed by adding a light quark field to the diquark composed of two heavy quarks. Besides, the interpolating current must also be a color singlet. Hence, the general form of the interpolating current for doubly heavy baryons can be written in the following form,
\begin{equation}
    J = \varepsilon^{abc} \left(Q^{aT} C \Gamma_1 Q^b \right) \Gamma_2 q^c~,
    \label{306}
\end{equation}
where $a$, $b$, $c$ are color indices and  $\Gamma_1$, $\Gamma_2$  are the Dirac matrices from the full set $\Gamma_i = \mathbb{1},\gamma_5,\gamma_\mu,\gamma_5\gamma_\mu,\sigma_{\mu\nu}$. Having found the form of the interpolating current, the next problem is the determination of $\Gamma_1$ and $\Gamma_2$ matrices. First, let us consider the diquark part of the interpolating current
\begin{equation}
    \varepsilon^{abc} \left(Q^{aT} C \Gamma_1 Q^b \right)~.
    \label{307}
\end{equation}
After taking the transpose of this equation, we get
\begin{equation}
    \left( \varepsilon^{abc}Q^{aT}C\Gamma_1Q^b\right)^T = -\varepsilon^{abc} Q^{bT} \Gamma_1^T C^T Q^a~,
    \label{308}
\end{equation}
where the fermion nature of the quark fields are taken into account.
Since $C^T = C^{-1} = C^\dagger = -C$,  we obtain
\begin{equation}
    \left(\varepsilon^{abc} Q^{aT} C \Gamma_1 Q^b \right)^T = \varepsilon^{abc} Q^{bT} C \left( C \Gamma_1^T C^{-1} \right) Q^a~.
    \label{309}
\end{equation}
Moreover, using the relation
\begin{equation}
    C\Gamma_1C^{-1} = 
    \begin{cases}
        \Gamma_1  &\text{for  } \Gamma_1 = {\mathbb{1},\gamma_5,\gamma_5\gamma_\mu}\\
        -\Gamma_1  &\text{for  } \Gamma_1 = {\gamma_\mu,\sigma_{\mu\nu}}\\
    \end{cases}
    \label{310}
\end{equation}
we get the following form for the diquark part of the interpolating current
\begin{equation}
\begin{aligned}
    \left(\varepsilon^{abc}Q^{aT}C\Gamma_1Q^b\right)^T = 
    \begin{cases}
        -\varepsilon^{abc}Q^{aT}C\Gamma_1Q^b &\text{for } \mathbb{1},\gamma_5,\gamma_5\gamma_\mu\\
        \varepsilon^{abc}Q^{aT}C\Gamma_1Q^b &\text{for } \gamma_\mu,\sigma_{\mu\nu}~.\\
    \end{cases}
        \label{311}
    \end{aligned}
\end{equation}
Since left side of this equation is represented by a $1 \times 1$ transposed matrix, it should be equal to itself. Hence, $\Gamma_1=\mathbb{1},\gamma_5 \text{ and } \gamma_5 \gamma_\mu$ structures are ruled out.

Second, let us consider the spin of the baryons. For the baryon in diquark quark picture a light quark with spin $S = \nicefrac{1}{2}$ and $S_z = \pm\nicefrac{1}{2}$ is attached to the diquark. Since we consider the doubly heavy baryon with $J=\nicefrac{1}{2}$, the diquark part should have zero spin. This implies that either $\Gamma_1 = \mathbb{1}$  or $ \Gamma_1 = \gamma_5$. This fact allows us to write the interpolating current in two possible forms
\begin{equation}
    \begin{aligned}
    j_1 &= \varepsilon^{abc} \left(Q^{aT} C q^b \right) \Gamma_3 Q^c \\
    j_2 &= \varepsilon^{abc} \left( Q^{aT} C \gamma_5 q^b \right) \Gamma_4 Q^c~.
    \end{aligned}
    \label{312}
\end{equation}
The form of the $\Gamma_3$ and $\Gamma_4$ can be determined by analyzing the Lorentz structures and parity consideration. The currents $j_1$ and $j_2$ must be a Lorentz scalar. This leads that $\Gamma_3$ and $\Gamma_4$ must be either $\mathbb{1} \text{ or }\gamma_5$.

To determine the final form of $\Gamma_3$ and $\Gamma_4$, let us consider the parity transformation. Applying the parity transformation to each quark field, we get
\begin{equation}
    \begin{aligned}
    j_1' &= \varepsilon^{abc}\left[\left(\gamma_0Q^a\right)^TC\gamma_0q^b\right]\Gamma_3\gamma_0Q^c  = \varepsilon^{abc}\left(Q^{aT}Cq^b\right)\Gamma_3\gamma_0Q^c \\
    j_2' &= \varepsilon^{abc}\left[\left(\gamma_0Q^a\right)^TC\gamma_0q^b\right]\Gamma_3\gamma_0Q^c  = \varepsilon^{abc}\left(Q^{aT}C\gamma_5q^b\right)\Gamma_4\gamma_0Q^c~.
    \end{aligned}
    \label{313}
\end{equation}
On the other side, applying the parity transformation to the interpolating currents, we obtain
\begin{equation}
    \begin{aligned}
        j_1' &= \varepsilon^{abc}\left(Q^{aT}Cq^b\right)\gamma_0\Gamma_3Q^c\\
        j_2' &= \varepsilon^{abc}\left(Q^{aT}C\gamma_5q^b\right)\gamma_0\Gamma_4Q^c~.
    \end{aligned}
    \label{314}
\end{equation}
Comparing Eqs. \eqref{313} and \eqref{314}, we find 
\begin{equation}
    \begin{aligned}
        -\Gamma_3\gamma_0 &= \gamma_0\Gamma_3\\
        \Gamma_4\gamma_0 &= \gamma_0\Gamma_4~.
    \end{aligned}
    \label{315}
\end{equation}
From these results, we get $\Gamma_3 = \gamma_5$ and $\Gamma_4 = \mathbb{1}$. Hence, we obtain the two possible forms of the interpolating current of double heavy baryon with spin-$\nicefrac{1}{2}$ 
\begin{equation}
    \begin{aligned}
        j_1 &= \varepsilon^{abc}\left(Q^{aT}Cq^b\right)\gamma_5Q^c\\
        j_2 &= \varepsilon^{abc}\left(Q^{aT}C\gamma_5q^b\right)Q^c~.
    \end{aligned}
    \label{316}
\end{equation}
The linear combination also describes the interpolating current of a spin-$\nicefrac{1}{2}$ doubly heavy baryon, i.e.
\begin{equation}
    J = 2\varepsilon^{abc}\left\{\left(Q^{aT}Cq^b\right)\gamma_5Q^c + \beta\left(Q^{aT}C\gamma_5q^b\right)Q^c\right\},
    \label{317}
\end{equation}
where $\beta$ is an arbitrary parameter.

By using the Fierz expansion, this current can be written as
\begin{equation}
    \begin{aligned}
        J = \frac{1-\beta}{2}\varepsilon^{abc}\left(Q^{aT}C\gamma_\mu Q^b\right)\gamma_5\gamma^\mu q^c 
        \notag + \frac{1+\beta}{4}\varepsilon^{abc}\left(Q^{aT}C\sigma_{\mu\nu}Q^b\right)\gamma_5\sigma^{\mu\nu}q^c~.
    \end{aligned}
    \label{318}
\end{equation}
The $\beta=-1$ case is called the Ioffe current in the literature. This form of the current was introduced by Ioffe in~\cite{Ioffe:1981kw} to study the mass of ordinary baryons within QCD sum rules. In order to construct the sum rules, the correlation function is calculated in two different regions, namely, in terms of hadrons (so-called hadronic side) and in terms of quark and gluons in deep Euclidean region (so-called theoretical part).

Having determined the form of the interpolating current for the baryon in diquark-quark picture, let us calculate the correlation function from the hadronic side.
The correlation function from the hadronic side can be calculated with the help of the dispersion relation, which connects it by its imaginary part
\begin{equation}
    \Pi(p^2) = -\int ds \frac{\rho(s)}{p^2-s+i\varepsilon} + \text{ ...}~,
    \label{319}
\end{equation}
where dots denote the possible subtraction terms. The term $\rho(s)$ is called the spectral density and is equal to the imaginary part of $\Pi$, i.e. $\rho(s) = \Im \Pi(s)$. The main reason for the dispersion relation being used instead of $\Pi$ itself is that calculations are easier.

Now let us turn our attention to the calculation of the correlation function from QCD side. The correlation function is calculated in deep Euclidean region with the help of operator product expansion (OPE). To calculate the spectral density, we follow the procedure below.
Let us consider the correlation function
\begin{equation}
    \Pi \left(p^2\right) = \int \frac{\rho(s)}{s-p^2}ds~.
    \label{327}
\end{equation}
As a first step, we take the Borel transformation of both sides with respect to the variable $p^2$, and we get
\begin{equation}
    \Pi^\mathcal{B} = \int \rho(s) e^{\frac{-s}{M^2}}ds~.
    \label{328}
\end{equation}
Denoting $\frac{1}{M^2}$ by $\tau$ and performing Borel transformation over the variable $\tau$ by using
\begin{equation}
    \lim_{\tau \to M^{-2}} \mathcal{B} e^{-s\tau} = \delta \left( \frac{1}{M^2} -s \right)
    \label{329}
\end{equation}
and integrating over $s$, we get $\rho = \tilde{\tilde{\Pi}}$, in which $\tilde{\tilde{\Pi}}$ means double Borel transformation.
This observation is essential for the calculation of spectral density, i.e. the spectral density $\rho$ can be obtained by applying double Borel transformation to the correlation function.
With this information, we present the spectral density calculation for one term as an example here. Using Eqs. \eqref{301} and \eqref{318} as well as Wick's theorem for the correlation function we get 

\begin{equation}
    \begin{split}
      \Pi\left(p^2\right) =&i\varepsilon^{abc}\varepsilon^{a'b'c'}\displaystyle\int d^4xe^{ipx} \times \\
      \bra{0}
        \Bigg\{ & - \gf \prop{cb'}{Q}\tilprop{ba'}{q}\prop{ac'}{Q}\gf 
        + \gf\prop{cc'}{Q}\gf\Tr [\prop{ab'}{Q}\tilprop{ba'}{q}]\\
        +&\beta \Bigg[ -\gf\prop{cb'}{Q}\gf\tilprop{ba}{q}\prop{ac'}{Q}
        -\prop{cb'}{Q}\tilprop{ba'}{q}\gf\prop{ac'}{Q} \gf\\
        +&\gf\prop{cc'}{Q} \Tr[\prop{ab'}{Q}\gf\prop{ab'}{Q}\gf\tilprop{ba'}{q}]
        +\prop{cc'}{Q}\gf \Tr\prop{ab'}{Q}\tilprop{ba'}{q} \gf] \Bigg]\\
        +&\beta^2\Bigg[-\prop{cb'}{Q}\gf\tilprop{ba'}{q}\gf\prop{ac'}{Q}
        + \prop{cc'}{Q}\Tr[\prop{ba'}{q}\gf\tilprop{ab'}{Q} \gf] \Bigg]\Bigg\}\ket{0}~,
    \end{split}
    \label{334}
\end{equation}
where $\widetilde{S} = CS^TC$.
The expressions of the light ($S_q$) and heavy quark ($S_Q$) propagators are needed to calculate the spectral density. These expressions are
\begin{equation}
    \begin{aligned}
        S_q = & \frac{i\slashed{x}}{2\pi^2x^4} - \frac{m_q}{4\pi^2x^2} - \frac{\expval{\Bar{q}q}}{12} \left( 1 - \frac{i m_q}{4}\slashed{x} \right) - \frac{x_0^2}{192} m_0^2 \expval{\Bar{q}q}\left( 1 - \frac{i m_q}{6} \slashed{x}\right)\\
        S_Q = & \frac{i m_Q^2}{\left( 2\pi \right)^2} \left[ i\slashed{x}\frac{K_2\left(m_Q\sqrt{-x^2} \right)}{\left( \sqrt{-x^2} \right)^2} + \frac{K_1\left( m_Q \sqrt{-x^2} \right)}{\sqrt{-x^2} }\right]~,
    \end{aligned}
    \label{335}
\end{equation}
where $K_i$ are the modified Bessel functions of the second kind.

Having the expressions for light and heavy quark propagators, we would like to present the details of the calculations of the spectral density for one of the term above and consider the following generic term as an example
\begin{equation}
    I = \displaystyle\int d^4x e^{ipx} \frac{1}{\left( x^2 \right)^{n_1}} \frac{K_{n_{2}}\left( m_Q \sqrt{-x^2} \right)}{\left( \sqrt{-x^2} \right)^{n_2}} \frac{K_{n_{3}}\left( m_Q \sqrt{-x^2} \right)}{\left( \sqrt{-x^2} \right)^{n_3}}~.
    \label{336}
\end{equation}
To perform the calculations, we need to transform from Minkowskian space to the Euclidean space, i.e. $-x^2 \rightarrow x_E^2$, $d^4x \rightarrow -d^4 x_E$, and $i p x \rightarrow -i p_E x_E$. Then using the Schwinger representation,
\begin{equation}
    \frac{1}{\left(x_E^2\right)^n} = \frac{1}{\Gamma(n)}\displaystyle\int_0^{\infty} dt \, t^{n-1} \, e^{-t_1x_E^2}
    \label{337}
\end{equation}
and integral representation for the Bessel function
\begin{equation}
    \frac{K_n\inp{m_Q\sqrt{-x^2}}}{\inp{\sqrt{-x^2}}^n} = \frac{2^{n-1}}{m_Q^n} \displaystyle\int_0^{\infty} dt_1 \, t_1^{n-1} \, e^{-x_E^2 t_1 - \frac{m_Q^2}{4t_1}}
    \label{338}
\end{equation}
for $I$, we get
\begin{equation}
    \begin{aligned}
        I =& \inp{-1}^n \frac{2^{n_2-1}}{\inp{m_Q}^{n_2}}\frac{2^{n_3-1}}{\inp{m_Q}^{n_3}} \displaystyle \int dt_1 \displaystyle \int dt_2 \displaystyle \int dt_3   t_1^{n_1-1}\, t_2^{n_2-1}\,t_3^{n_3-1}\\
        &\displaystyle \int d^4 x_E   e^{-\inp{t_1+t_2+t_3}x_E^2 - \frac{m_Q^2}{4t_2} - \frac{m_Q}{4t_3} - ip_E x_E}~.
    \end{aligned}
    \label{339}
\end{equation}
Integrating over $d^4 x_E $, we get
\begin{equation}
    \begin{aligned}
        I =& A i \pi^2 \displaystyle\int dt_1\,dt_2\,dt_3 t_1^{n_1-1}\,t_2^{n_2-1}\,t_3^{n_3-1}\\
        &\frac{1}{\inp{t_1+t_2+t_3}^2} e^{\frac{-p_E^2}{4\inp{t_1+t_2+t_3}}-\frac{m_Q^2}{4t_2}-\frac{m_Q^2}{4t_3}}~,
    \end{aligned}
    \label{340}
\end{equation}
where $\displaystyle A = \inp{-1}^{n_1} \frac{1}{\Gamma\inp{n_1}}\frac{2^{n_2+n_3-2}}{\inp{m_Q}^{n_2+n_3}}$~.

Now let us introduce new variables
\begin{equation}
    \begin{aligned}
        t_1 & = t \inp{1-x-y}\\
        t_2 & = t x\\
        t_3 & = t y~.
    \end{aligned}
    \label{341}
\end{equation}
Then $dt_1 dt_2 dt_3\rightarrow dx dy dt t^2$. With this transformation, we get
\begin{equation}
    \begin{aligned}
        I = A i \pi^2 \displaystyle \int& dx\,dy\,dt\,e^{-\frac{p_E^2}{4t} - \frac{m_Q^2}{4tx} - \frac{m_Q^2}{4ty}}
         t^{n_1+n_2+n_3-3} x^{n_2-1} y^{n_3-1} \inp{1-x-y}^{n_1-1}~.
    \end{aligned}
    \label{342}
\end{equation}
Once we denote $\displaystyle \frac{1}{4t}\inb{p_E^2+\frac{m_Q^2}{x}+\frac{m_Q^2}{y}} = \frac{\alpha}{4 t } = t'$, with $\displaystyle \alpha = p_E^2+\frac{m_Q^2}{x}+\frac{m_Q^2}{y}$, then $\displaystyle dt = \frac{dt'}{4t^{'2}}\alpha$ and Eq. \eqref{340} turns out to be
\begin{equation}
    \int dt' \frac{1}{4t^{'2}} \alpha \inp{\frac{\alpha}{4t^{'2}}^{\sum n_i - 3}} e^{-t'} = \frac{\alpha^{\sum n_i - 2}}{4^{\sum n_i - 2}} \Gamma \inp{2 - \sum n_i}~.
    \label{343}
\end{equation}
As a result, we get
\begin{equation}
    I = A i \pi^2 \frac{\Gamma\inp{2 - \sum n_i}}{4^{\sum n_i - 2}} \int dx\,dy x^{n_2-1}y^{n_3-1}\inp{1-x-y}^{n_1-1} \frac{1}{\alpha^{2 - \sum n_i}}~.
    \label{344}
\end{equation}
For $\alpha^{2 - \sum n_i}$, we use the Schwinger representation once more.
\begin{equation}
    \frac{1}{\alpha^{2 - \sum n_i}} = \frac{1}{\Gamma\inp{\alpha^{2 - \sum n_i}}}\int dt  t^{1 - \sum n_i} e^{-t\inb{\frac{m_Q^2}{x}+\frac{m_Q^2}{y}+p_E^2}}
    \label{345}
\end{equation}
Now performing Borel transformation with respect to the variable $p_E^2$, i.e. $\displaystyle e^{-t p_E^2} \rightarrow \delta\inp{\sigma - t}$, where $\sigma = 1/M^2$ and after integrating over $t$ we get
\begin{equation}
    \begin{aligned}
        I^\mathcal{B} = & A i \pi^2 \frac{1}{4^{\sum n_i -2}} \int dx\,dy
         x^{n_2-1} y^{n_3-1}\inp{\sigma}^{1 - \sum n_i} e^{ - \sigma \inp{\frac{m_Q^2}{x} + \frac{m_Q^2}{y}}}~.
    \end{aligned}
    \label{346}
\end{equation}
Using the Schwinger representation for $\displaystyle \frac{1}{\sigma^{\sum n_i -1}}$ for $I^B$, we get
\begin{equation}
    \begin{aligned}
        I^\mathcal{B} = \frac{Ai\pi^2}{4^{\sum n_i -2}} &\frac{1}{\Gamma\inp{\sum n_i -1}} \int dx\int dy\int d\alpha   \alpha^{\sum n_i -2}
         e^{-\alpha\sigma} e^{-\sigma \inb{\frac{m_Q^2}{x}+\frac{m_Q^2}{y}}}~.
    \end{aligned}
    \label{347}
\end{equation}
After performing the second Borel transformation with respect to the variable $\sigma$, we get
\begin{equation}
    \begin{aligned}
        I^\mathcal{BB} =& \frac{Ai\pi^2}{4^{\sum n_i -2}} \frac{1}{\Gamma\inp{\sum n_i -1}} \int dx\int dy\int d\alpha   \alpha^{\sum n_i -2} \delta \inp{s - \alpha - \frac{m_Q^2}{x} - \frac{m_Q^2}{y}}\\
        =& \frac{Ai\pi^2}{4^{\sum n_i -2}} \frac{1}{\Gamma\inp{\sum n_i -1}} \int dx\int dy \inp{s - \frac{m_Q^2}{x} - \frac{m_Q^2}{y}}^{\sum n_i -2}~.
    \end{aligned}
    \label{348}
\end{equation}
This expression is the spectral density for $I$. The bounds of $x$ and $y$ variables are
\begin{equation}
    \begin{aligned}
      y & \geq \frac{m_Q^2}{s - \nicefrac{m_Q^2}{x}}
      s - \frac{m_Q^2}{x} - \frac{m_Q^2}{y} & \geq 0~.
    \end{aligned}
    \label{349}
\end{equation}
Since the upper bound of $y$ is the $1-x$, then the bounds of $y$ become
\begin{equation}
    \frac{m_Q^2}{s - \nicefrac{m_Q^2}{x}} \leq y \leq 1-x~.
    \label{350}
\end{equation}
On the other hand, the bounds of $x$ can be obtained from 
\begin{equation}
    \frac{m_Q^2}{s - \nicefrac{m_Q^2}{x}} \leq 1 - x
    \label{351}
\end{equation}
which leads to 
\begin{equation}
    x = \frac{s \pm \sqrt{s^2-4s\,m_Q^2}}{2s} = \frac{\inp{1\pm\slashed{v}}}{2},
    \label{352}
\end{equation}
where $\displaystyle v = \sqrt{1-4m_Q^2/s}$ is the velocity of the heavy quark.

Now we present the expression of the correlation function from the hadronic side. Inserting the full set of states carrying the same quantum numbers as the hadron under consideration we get
\begin{equation}
  \label{352a}
  \Pi = \sum_i \frac{(\slashed{p} + m_i)}{m_i^2 - p^2},
\end{equation}
where summation is performed over full set of states. From here, it follows that the interpolating current used in Eq.~\eqref{317} interacts not only with the ground state but also with the excited states carrying the same quantum numbers. In many cases, while the ground state is narrow, excited states are broader. Therefore, the spectral density can be parameterized as a single pole of the ground state plus a smooth continuum corresponding to the contributions of the higher states, i.e.
\begin{equation}
    \rho(s)=\lambda^2\delta(s-m^2)+\rho^{\text{cont}}(s)~.
    \label{320}
\end{equation}
In the above equation, $\lambda$ is called the decay constant (residue), and it is defined as 
\begin{equation}
    \lambda u(p)= \bra{0} j \ket{\text{ground state baryon}}~.
    \label{321}
\end{equation}
The contribution of the higher states is determined with the help of the quark-hadron duality ansatz, which means that the $\rho^{\text{cont}}$ coincides with the OPE part of the spectral density starting from some effective threshold $s_0$. In other words, the spectral density for the continuum can be written as
\begin{equation}
    \rho^{cont}(s) = \rho^{OPE} \theta \left(s-s_0\right)~.
    \label{330}
\end{equation}
As we already noted, the mass sum rules are obtained by matching the two representations of the correlation function.
\begin{equation}
    \Pi^{OPE} = \Pi^{had}
    \label{331}
\end{equation}
To enhance the contribution of the ground state as well as suppress the contributions of higher states and higher dimension operators from OPE part, Borel transformation is applied to both sides of Eq. \eqref{331}. After the Borel transformation, all the polynomial terms disappear.
The Borel transformation for the  expression $\frac{1}{\left(m^2-p^2\right)^n}$ is given by
\begin{equation}
   \frac{1}{\left(m^2-p^2\right)^n} \xrightarrow{\mathcal{B}_{M^2}}  \frac{1}{\left(n-1\right)!\left(M^2\right)^{n-1}}e^{\frac{-m^2}{M^2}} \,\, (\text{ n$>$0 })~.
    \label{332}
\end{equation}
After performing the Borel transformation in both representations of the correlation function, we get
\begin{equation}
    \lambda^2e^{\displaystyle\nicefrac{-m^2}{M^2}} = \displaystyle\int_{s_{min}}^{s_0}ds e^{\displaystyle\nicefrac{-s}{M^2}}\rho(s)~.
    \label{333}
\end{equation}
To derive this expression, the quark-hadron duality ansatz is employed.

The mass of low lying state hadron can be obtained by taking derivatives with respect to $\frac{-1}{M^2}$ from Eq. \eqref{333} and dividing the result to Eq. \eqref{333}. After this operation, we get the following expression for the ground state mass  
\begin{equation}
    m^2 = \frac{ \displaystyle \int_{s_{min}}^{s_0}\rho(s) s e^{\frac{-s}{M^2}}ds}{\displaystyle \int_{s_{min}}^{s_0} \rho(s) e^{\frac{-s}{M^2}}ds}~.
    \label{326}
\end{equation}
Performing the similar calculations for all the remaining terms, one can obtain the expressions of the spectral density (for the expressions of the spectral density, see \cite{Aliev:2012ru}.)
Using the explicit expression of the spectral density from Eq. \eqref{334}, one can determine the mass of the ground state baryon.

In addition to the input parameters such as heavy quark mass and various condensates, the mass sum rules contain three auxiliary parameters: Borel mass parameter $M^2$, continuum threshold $s_0$, and auxiliary parameter $\beta$. Any physically measurable quantity should be independent of these parameters. Therefore it is necessary to find so-called ``working regions" of these parameters, where the physical quantity (in our case mass) exhibits very weak dependency to them.

The working region of Borel mass is determined in the following way. The upper bound of $M^2$ is determined by the fact that the pole contribution should be larger than the continuum and the higher states contributions. Therefore, we demand that the ratio
\begin{equation}
    \text{P.C.} = \frac{\displaystyle\int_{s_{min}}^{s_0} \rho(s) \, e^{-s/M^2 }ds}{\displaystyle\int_{s_{min}}^{\infty} \rho(s) \, e^{-s/M^2}ds } 
    \label{353}
\end{equation}
should be larger than $1/2$.

The lower bound of $M^2$ is obtained from the condition that the OPE series should be convergent, i.e. the perturbative contribution should be larger than the nonperturbative one. Besides, the continuum threshold is determined from the condition that the differentiated sum rules reproduce the measured mass of the lowest state hadron with, say, $10\%$ accuracy.
Having the working regions of $M^2$ and $s_0$ with the conditions mentioned, we are left with to determine the parameter $\beta$. Analyzing the dependency of the mass on $\beta$, the working region where the mass is almost independent of the variation is obtained.  

A similar analysis was conducted for spin-$3/2$ doubly heavy baryons in~\cite{Aliev:2012iv}. The main difference, in this case, is the form of the interpolating current. In the quark-diquark picture, the diquark field has spin $1$.
After performing the numerical analysis, the obtained values for the mass and residues of doubly heavy baryons with spin-$1/2$ and spin-$3/2$ are collected in Tables \ref{tab:2} and \ref{tab:3}, respectively. For completeness, in these tables, we present the predictions for the mass and residues obtained from the other approaches.

\begin{table*}[t]
  \centering
  \renewcommand{\arraystretch}{1.4}
  \setlength{\tabcolsep}{3.2pt}
  \begin{tabular}{lcccccccc}
    \toprule
        Baryon & \cite{Aliev:2012ru} & \cite{Wang:2017mqp} & \cite{Zhao:2018mrg} & \cite{Xing:2018lre} & \cite{Shi:2019hbf} & \cite{Shi:2019fph} & \cite{Hu:2019bqj} & $\lambda$ \cite{Aliev:2012ru} \\
    \midrule
        \tablexi{}{bb} & 9.96 & 9.78 & 10.17 & 9.94 & 10.202 & - & - & 0.44 \\
        \tableom{}{bb} & 9.97 & 9.85 & 10.32 & 9.99 & 10.359 & - & - & 0.45 \\
        \midrule
        \tablexi{}{bc} & 6.72 & 6.75 & - & 6.80 & 6.933 & 7.053 & - & 0.28 \\
        \tableom{}{bc} & 6.75 & 7.02 & - & 6.864 & 7.088 & 7.148 & - & 0.29 \\
        \midrule
        \tablexi{}{cc} & 3.72 & 4.26 & 3.57 & 3.52 & 3.620 & 3.676 & 3.5189 & 0.16 \\
        \tableom{}{cc} & 3.73 & 4.25 & 3.71 & 3.53 & 3.778 & 3.787 & - & 0.18 \\
        \midrule
        \tablexi{'}{bc} & 6.79 & 6.95 & - & - & 6.963 & 7.062 & - & 0.30 \\
        \tableom{'}{bc} & 6.80 & 7.02 & - & - & 7.116 & 7.151 & - & 0.31 \\
    \bottomrule
  \end{tabular}
  \caption{The mass and residue of doubly heavy baryons with spin $\nicefrac{1}{2}$ in units GeV.}
  \label{tab:2}
\end{table*}

\begin{table*}[t]
  \centering
  \renewcommand{\arraystretch}{1.4}
  \setlength{\tabcolsep}{3.2pt}
  \begin{tabular}{lcccccc}
    \toprule
        Baryon          & $\slashed{q} g_{\mu\nu}$ \cite{Aliev:2012iv} & $g_{\mu\nu}$ \cite{Aliev:2012iv}    & \cite{Wang:2010vn} & \cite{Ebert:2002ig} & \cite{Zhang:2008rt} & \cite{Bagan:1992za,Narison:2010py} \\ 
    \midrule
        \tablexi{*}{cc} & $3.69 \pm 0.16$          & $3.72 \pm 0.18$ & $3.61 \pm 0.18$    & $3.727$             & $3.90 \pm 0.10$     & $3.58 \pm 0.05$                    \\
        \tableom{*}{cc} & $3.78 \pm 0.16$          & $3.78 \pm 0.16$ & $3.76 \pm 0.17$    & $3.872$             & $3.81 \pm 0.06$     & $3.67 \pm 0.05$                    \\
        \midrule
        \tablexi{*}{bb} & $10.4 \pm 1.0$           & $10.3 \pm 0.2$  & $10.22 \pm 0.15$   & $10.237$            & $10.35 \pm 0.08$    & $10.33 \pm 1.09$                   \\
        \tableom{*}{bb} & $10.5 \pm 0.2$           & $10.4 \pm 0.2$  & ---                & $6.98$              & $8.00 \pm 0.26$     & ---                                \\
        \midrule
        \tablexi{*}{bc} & $7.25 \pm 0.20$          & $7.2 \pm 0.2$   & ---                & $7.13$              & $7.54 \pm 0.08$     & ---                                \\
        \tableom{*}{bc} & $7.3 \pm 0.2$            & $7.35 \pm 0.25$ & $10.38 \pm 0.14$   & $10.389$            & $10.28 \pm 0.05$    & $10.38 \pm 1.10$                   \\
    \bottomrule
  \end{tabular}
  \caption{The mass spectra of the spin $3/2$ doubly heavy baryons in units of $\rm{GeV.}
    $~\cite{Aliev:2012iv}.}
  \label{tab:3}
\end{table*}

\begin{table*}[t]
  \centering
  \renewcommand{\arraystretch}{1.4}
  \setlength{\tabcolsep}{3.2pt}
  \begin{tabular}{lcccccc}
    \toprule
        Baryon          & $\slashed{q} g_{\mu\nu}$ \cite{Aliev:2012iv} & $g_{\mu\nu}$ \cite{Aliev:2012iv} & \cite{Wang:2010vn} & \cite{Bagan:1992za} \\ 
    \midrule
        \tablexi{*}{cc} & $0.12 \pm 0.01$                              & $0.12 \pm 0.01$                  & $0.070 \pm 0.017$  & $ 0.071 \pm 0.017$  \\
        \tableom{*}{cc} & $0.14 \pm 0.02$                              & $0.13 \pm 0.01$                  & $0.085 \pm 0.019$  & ---                 \\
        \midrule 
        \tablexi{*}{bb} & $0.22 \pm 0.03$                              & $0.21 \pm 0.01$                  & $0.161 \pm 0.041$  & $0.111 \pm 0.040$   \\
        \tableom{*}{bb} & $0.25 \pm 0.03$                              & $0.25 \pm 0.02$                  & ---                & ---                 \\ 
        \midrule
        \tablexi{*}{bc} & $0.15 \pm 0.01$                              & $0.15 \pm 0.01$                  & ---                & ---                 \\
        \tableom{*}{bc} & $0.18 \pm 0.02$                              & $0.17 \pm 0.01$                  & $0.199 \pm 0.048$  & ---                 \\
    \bottomrule
  \end{tabular}
  \caption{The residues of the spin $3/2$ doubly heavy baryons in units of $\rm{GeV}^3$~\cite{Aliev:2012iv}.}
  \label{tab:3a}
\end{table*}

The sum rules method allows not only calculating the properties of the ground state hadron but also the properties of radial or orbital excitations of these baryons. The modification of the sum rules is required only in the phenomenological part. Namely, Eq. \eqref{320} is modified as
\begin{equation}
    \rho(s) = \lambda_0^2 \delta\left(s-m_0^2\right) + \lambda_1^2 \delta\left(s-m_1^2\right),
    \label{354}
\end{equation}
where $m_0\left(m_1\right)$ and $\lambda_0\left(\lambda_1\right)$ are the mass and residues of the ground (excited) state, respectively.

The numerical calculations are performed likewise for the ground state baryons, which are explained in detail above and the results for $2S$ and $1P$ excited state baryons are summarized in Table~\ref{tab:4}.  In these tables, we also presented the results of other approaches on the mass of doubly heavy baryons. In Table~\ref{tab:4}, we also present the values of residues of $2S$ and $1P$ states. The obtained results on the mass and residues can be useful in looking for these states in future experiments.

\begin{table*}[t]
  \centering
  \renewcommand{\arraystretch}{1.4}
  \setlength{\tabcolsep}{3.2pt}
  \begin{tabular}{lcccc}
    \toprule
    Baryon & $\sqrt{s_0} (GeV)$ & $M^2 (GeV^2)$ & $2S (GeV)$ & $1P (GeV)$\\
    \midrule
        \tablexi{}{cc} & $4.3 \pm 0.1$ & $\inb{4-7}$ & $4.03 \pm 0.20$ & $4.03 \pm 0.20$ \\
        \tableom{}{cc} & $4.5 \pm 0.1$ & $\inb{4-7}$ & $4.16 \pm 0.14$ & $4.16 \pm 0.14$ \\
        \hline  
        \tablexi{}{bb} & $11.1 \pm 0.1$ & $\inb{10-16}$ & $10.32 \pm 0.10$ & $10.32 \pm 0.10$ \\
        \tableom{}{bb} & $11.2 \pm 0.1$ & $\inb{10-16}$ & $10.37 \pm 0.10$ & $10.32 \pm 0.10$ \\
        \hline
        \tablexi{}{bc} & $7.7 \pm 0.1$ & $\inb{7-11}$ & $7.14 \pm 0.11$ & $7.14 \pm 0.11$ \\
        \tableom{}{bc} & $7.9 \pm 0.1$ & $\inb{7-11}$ & $7.20 \pm 0.11$ & $7.20 \pm 0.11$ \\
        \hline 
        \tablexi{'}{bc} & $7.9 \pm 0.1$ & $\inb{7-11}$ & $7.02 \pm 0.07$ & $7.02 \pm 0.07$ \\
        \tableom{'}{bc} & $7.9 \pm 0.1$ & $\inb{7-11}$ & $7.09 \pm 0.07$ & $7.09 \pm 0.07$ \\
    \bottomrule
  \end{tabular}
  \caption{The working domains of continuum threshold and Borel parameters $M^2$ along with the respective mass of $2S$ and $1P$ excitations of the doubly heavy baryons \cite{Aliev:2019lvd}.}
  \label{tab:4}
\end{table*}

\section{Calculation of the strong coupling constants of doubly heavy baryons with light mesons}
\label{sec:4}
As we already noted, for a deeper understanding of the dynamics of the double heavy baryons, the investigation of the weak, strong, and electromagnetic decays of the baryons should be conducted. Up to now, the studies of the strong decays have been very limited. Hence, detailed studies of strong decays can receive great interest for the study of the properties of these baryons.

In this part of the mini-review, we study the strong coupling constants of doubly heavy baryons with light pseudo-scalar and vector mesons within the light-cone version of sum rules. Moreover, by using these coupling constants and applying the vector meson dominance model, we estimate the radiative decays of double heavy baryons.

To determine the strong coupling constants of the doubly heavy baryons with light mesons within light-cone sum rules, we start with considering the following correlation function.
\begin{equation}
    \Pi = i\int d^4x   e^{ipx} \bra{M(q)}\eta(x)\,\bar{\eta}(0)\ket{0},
    \label{355}
\end{equation}
where $M(q)$ is a meson with momentum $q$, $\eta$ is the interpolating current of the corresponding doubly heavy baryon. According to the $SU(3)$ classification, there are two types of currents, symmetric and antisymmetric currents, with respect to the heavy quarks. The interpolating currents (symmetric and antisymmetric) for doubly heavy baryons with $ J = \nicefrac{1}{2}$ are
\begin{equation}
    \begin{aligned}
        \eta^{(S)} =& \frac{1}{\sqrt{2}}\varepsilon^{abc} \sum_{i=1}^2\left\{\left(Q^{aT}A^i_1q^b\right)A^i_2Q'^c + \left(Q \xleftrightarrow{} Q'\right)\right\} \\
        \eta^{(A)} =& \frac{1}{\sqrt{6}}\varepsilon^{abc} \sum_{i=1}^2\Big\{2\left(Q^{aT}A^i_1Q'^b\right)A^i_2q^c
        + \left(Q^{aT}A^i_1q^b\right)A^i_2Q'^c - \left(Q'^{aT}A^i_1q^b\right)A^i_2Q^c \Big\},
    \end{aligned}
    \label{356}
\end{equation}
where $A_{1}^{1} = C$, $A_{1}^{2} = C\gf$ , $A_{2}^{1} = \gf$ and $A_{2}^{2} = \beta \mathbb{1}$ and $S$ and ${A}$ corresponds to symmetric and antisymmetric currents.

Similar to the mass sum rules calculation, the correlation function is calculated in two different regions, namely, in hadrons and the deep Euclidean domain $p^{2} \ll 0$ and $\left(p+q\right)^{2} \ll 0$ by using the OPE which is carried out over the twist rather than dimensions of operators. Then, performing Borel transformation in order to suppress the continuum contributions and enhance the contribution of the ground state and matching these two results, we get the desired sum rules.

The representation of the correlation function in terms of the hadron is obtained by inserting a complete set of baryon states having the same quantum numbers as the interpolating current. Then, isolating the contribution of the ground state, we get
\begin{equation}
    \Pi = \frac{\bra{0}\eta\ket{B(p')} \bra{M(q)B(p')}\ket{B(p)}\bra{B(p)}\bar{\eta}\ket{0}}{\left(p'^2-m_2^2\right)\left(p^2-m_1^2\right)} \hfill +\text{higher states}~.
    \label{357}
\end{equation}
The matrix elements entering to the Eq. \eqref{357} for the pseudoscalar meson case are determined as
\begin{equation}
    \bra{0}\eta\ket{B(p')} = \lambda_2 u\left(p'\right)
    \label{358}
\end{equation}
\begin{equation}
    \bra{M(q)B(p')}\ket{B(p)} = g\bar{u}\left(p'\right) i\gf u\big(p\big)~.
    \label{359}
\end{equation}
For the vector meson case, the matrix element $\bra{M(q)B(p')}\ket{B(p)}$ is defined as 
\begin{equation}
    \bra{M(q)B(p')}\ket{B(p)} = \bar{u}(p')\left[f_1\gamma_\mu + \frac{i\sigma_{\mu\nu}q^\nu}{m_1+m_2}f_2\right] u(p)\varepsilon^\mu~,
    \label{360}
\end{equation}
where $f_1$ and $f_2$ are the relevant form factors of the doubly heavy baryons with corresponding light vector mesons and $\varepsilon^\mu$ is the polarization vector of meson.

Using Eqs.~\eqref{357} and \eqref{360} and after performing double Borel transformations over $-p^2$ and $-p'^2$ for the correlation functions containing pseudo-scalar and vector mesons,
\begin{equation}
    \Pi^{(P)} = \lambda_1\lambda_2ge^{\frac{-(m_{B_1}^2 + m_{B_2}^2)}{2M^2}}  \hfill ( \text{for } \slashed{p}\slashed{q}\gf \text{ structure + ...})
    \label{361}
\end{equation}
\begin{equation}
    \Pi^{(V)} = \lambda_1\lambda_2\left[\slashed{p}\slashed{\varepsilon}\slashed{q}(f_1+f_2) + 2(\varepsilon p)\slashed{p}f_1 + \text{ ...}\right] e^{\frac{-(m_{B_1}^2 + m_{B_2}^2)}{2M^2}} ~,
    \label{362}
\end{equation}
where dots denote other structures.

Calculating the correlation functions from the QCD side and separating the structures $\slashed{p}\slashed{q}\gf$, $\slashed{p}\slashed{\varepsilon}\slashed{q}$ and $2(\varepsilon p)\slashed{p}$ and equating them to the corresponding coefficients of the same structures, one can obtain the sum rules for the strong coupling constants for the pseudo-scalar and vector mesons. The new and essential element for the calculation of the theoretical part of the correlation function is the so-called distribution amplitudes (DA) of light mesons. 

The DAs contain all information about the nonperturbative sector of QCD and are obtained in many works~\cite{Ball:1998ff,Ball:2006wn,Ball:2004ye,Ball:1998sk,Ball:2002ps}.
The main problem is the calculation of the spectral density to determine the theoretical part of the correlation function. We here present the details of the calculation for the generic term.
\begin{equation}
    I = \int_0^1du\frac{u^k}{\left[p_1^2\bar{u}+p_2^2u+m_M^2\bar{u}u+m^2\right]^n}~,
    \label{363}
\end{equation}
where $m_M$ is the mass of the hadron or photon (which is zero), and all momenta are represented in Euclidean space. Using the Schwinger representation, we get
\begin{equation}
    I = \int_0^1du\int_0^\infty d\alpha \alpha^{n-1}u^k\ e^{-\alpha\left[p_1^2\bar{u}+p_2^2u+m_M^2\bar{u}u+m^2\right]}~.
    \label{364}
\end{equation}
Performing double Borel transformation over $p_1\rightarrow\frac{1}{\sigma_1}$ and $p_2\rightarrow\frac{1}{\sigma_2}$ we have
\begin{equation}
    \begin{aligned}
        I^{\mathcal{B}} =& \int_0^1\int_0^\infty \alpha^{n-1} u^k e^{-\alpha(m_M^2\bar{u}u+m^2)}\delta(\sigma_1 - \alpha\bar{u})\delta(\sigma_2-\alpha u)\\
        =& (\sigma_1+\sigma_2)^{n-k-2}\sigma^k e^{(-\sigma_1+\sigma_2)}\left(m^2+\frac{m_M^2}{2}\right) + m_M^2\frac{\sigma_1^2+\sigma_2^2}{2(\sigma_1+\sigma_2)}~.
    \end{aligned}
    \label{365}
\end{equation}
Once we use the integral representation
\begin{equation}
    \int_{-\infty}^\infty e^{-\rho x^2 - \beta x}  = \displaystyle\sqrt{\frac{\pi}{\alpha}} e^{\frac{-\beta^2}{4\rho}}~.
    \label{366}
\end{equation}
Eq.~\eqref{365} can be written as
\begin{equation}
    \begin{aligned}
        I^\mathcal{B} = \int_{-\infty}^\infty\frac{d\sigma_1 d\sigma_2}{2\pi}(\sigma_1+\sigma_2)^{n-k-1}\sigma_2^k 
        e^{-\sigma_1\left(m^2+\frac{(m_M+x_1)^2 + x_2^2}{2}\right) -\sigma_2\left(m^2+\frac{(m_M+x_2)^2 + x_1^2}{2}\right)}~.
    \end{aligned}
    \label{367}
\end{equation}
Again using the Schwinger representation for $\frac{1}{\left(\sigma_1+\sigma_2\right)^{n-k-1}}$, we get the following expression for the spectral density 
\begin{equation}
    \rho(s_1,s_2) = \left(\frac{d}{ds_2}\right)^k\frac{1}{2m_M\pi}\frac{\Gamma(k)}{\Gamma(2k)}\left(\frac{\sqrt{\Delta}}{m_M}\right)^{2k-1}\theta(\Delta)~,
    \label{368}
\end{equation}
where
\begin{equation}
    \begin{aligned}
        \Delta&=\left[-m_M^4 - (s_1 - s_2)^2 + 2m_M^2(-2m^2 + s_1 + s_2)\right]\\
        &=(t_2 - s_2)(s_1 - t_1)~.
    \end{aligned}
    \label{369}
\end{equation}
Omitting the details of the calculations, we present the numerical results for the strong coupling constants of the light mesons with heavy baryons in Table~\ref{tab:5}. The interested reader can find the details of the calculations for sum rules in~\cite{Aliev:2020aon,Alrebdi:2020rev}. 

\begin{table*}[t]
  \centering
  \renewcommand{\arraystretch}{1.4}
  \setlength{\tabcolsep}{3.2pt}
  \begin{tabular}{lcc}
    \toprule
         & Channel & Strong coupling constant\\
    \midrule
             & \tablexi{}{cc}\tablexi{}{cc}$\pi$ & $10.3  \pm 0.52$ \\
             & \tablexi{}{bb}\tablexi{}{bb}$\pi$ & $12.73 \pm 1.29$ \\
        $SS$ & \tableom{}{bb}\tablexi{}{bb}$K$   & $17.40 \pm 1.89$ \\
             & \tableom{}{cc}\tablexi{}{cc}$K$   & $12.50 \pm 0.75$ \\
             & \tableom{}{bc}\tablexi{}{bc}$K$   & $5.08 \pm 0.43$ \\
        \hline
        \multirow{2}{*}{$AA$} & \tablexi{'}{bc}\tablexi{'}{bc}$\pi$ & $6.85 \pm 0.06$ \\
                              & \tableom{'}{bc}\tablexi{'}{bc}$K$   & $7.90 \pm 0.16$ \\
        \hline
        \multirow{2}{*}{$SA$} & \tablexi{'}{bc}\tablexi{}{bc}$\pi$ & $1.49 \pm 0.10$ \\
                              & \tableom{'}{bc}\tablexi{}{bc}$K$   & $2.03 \pm 0.16$ \\
    \bottomrule
  \end{tabular}
  \caption{The numerical values for the strong coupling constants \cite{Alrebdi:2020rev}.}
  \label{tab:5}
\end{table*}

\begin{table*}[t]
  \centering
  \renewcommand{\arraystretch}{1.4}
  \setlength{\tabcolsep}{3.2pt}
  \begin{tabular}{lccc}
    \toprule
         & Channel & $f_1  + f_2$ & $f_1$ \\
    \midrule
             & \tablexi{}{cc}$\rightarrow$\tablexi{}{cc}$\rho$ & $32.53 \pm 2.30$ & $-25.32 \pm 8.49$ \\
             & \tablexi{}{bb}$\rightarrow$\tablexi{}{bb}$\rho$ & $23.66 \pm 2.89$ & $-7.69 \pm 2.31$ \\ 
        $SS$ & \tableom{}{bb}$\rightarrow$\tablexi{}{bb}$K^{*}$ & $22.55 \pm 2.63$ & $8.30 \pm 2.74$ \\
             & \tableom{}{cc}$\rightarrow$\tablexi{}{cc}$K^{*}$ & $28.36 \pm 1.83$ & $-23.64 \pm 9.85$ \\
             & \tableom{}{bc}$\rightarrow$\tablexi{}{bc}$K^{*}$ & $8.23 \pm 0.77$ & $-3.74 \pm 1.37$ \\ 
        \hline
        \multirow{2}{*}{$AA$}   & \tablexi{'}{bc}$\rightarrow$\tablexi{'}{bc}$\rho$ & $-37.62 \pm 0.48$ & $-0.40 \pm 0.12$ \\
                                & \tableom{'}{bc}$\rightarrow$\tablexi{'}{bc}$K^{*}$ & $-37.98 \pm 6.75$ & $-0.40 \pm 0.14$\\
        \hline                        
        \multirow{2}{*}{$SA$}   & \tablexi{'}{bc}$\rightarrow$\tablexi{}{bc}$\rho$ & $1.50 \pm 0.31$ & $-0.97 \pm 0.30$ \\
                                & \tableom{'}{bc}$\rightarrow$\tablexi{}{bc}$K^{*}$ & $2.20 \pm 0.40$ & $-1.00 \pm 0.36$ \\
    \bottomrule
  \end{tabular}
  \caption{The numerical values for the strong coupling constants \cite{Aliev:2020aon}.}
  \label{tab:6}
\end{table*}

Now let us calculate the strong coupling constants of the spin-$\nicefrac{3}{2}$-spin$\nicefrac{1}{2}$ doubly heavy baryons with light vector mesons.
The calculation scheme is similar to the calculation of spin$\nicefrac{1}{2}$-spin$\nicefrac{1}{2}$ vector mesons. However, in this case, some difficulties appear for such types of transitions.

In spin-$\nicefrac{3}{2}-$spin-$\nicefrac{1}{2}$ vector strong couplings, we need the interpolating currents of spin$\nicefrac{3}{2}$ particles first. Following the same procedure for constructing interpolating current for spin$\nicefrac{1}{2}$ doubly heavy baryon states, one can easily obtain the following expressions for the interpolating currents for spin$\nicefrac{3}{2}$ baryons
\begin{equation}
    \eta_{\mu} = N \, \varepsilon^{abc} \left\{ \inp{q^{aT} C \gamma_{\mu} Q^b } Q^{'C} + \inp{q^{aT} C\gamma_{\mu} Q^{'b} } Q^c + \inp{Q^{aT} C \gamma_{\mu} Q^{'b}} q^c \right\}~,
\end{equation}
where $\displaystyle N = \sqrt{1/3}\inp{\sqrt{2/3}}$ for identical (different) heavy quarks.
The following matrix elements appear for the calculation of the phenomenological part, 
\begin{equation}
    \begin{aligned}
        \bra{0} \eta_{\mu} \ket{B^{*}\inp{p}} = & \lambda \, u_{\mu} \inp{p}\\
        \bra{V\inp{q} B^{*}\inp{p_2}} \ket{B\inp{p_1}} = & \bar{u}_{\alpha} \inp{p_2} \bigg[g_1 \inp{\varepsilon_{\alpha}\slashed{q} - q_{\alpha}\slashed{\varepsilon}} \gamma_5  
          + g_2 \inp{P q \varepsilon_{\alpha} - P\varepsilon q_{\alpha}} \gamma_5 
          + g_3 \inp{q\varepsilon q_{\alpha} - q^2\varepsilon_{\alpha}}\gamma_5 \bigg] u\inp{p_1}, 
    \end{aligned}
    \label{370}
\end{equation}
where $u_{\alpha}\inp{u}$ is the Rarita--Schwinger (Dirac) spinor, $\varepsilon_{\mu}$ is the polarization vector of the light vector meson, $P = (p_1 + p_2)$, and $q = p_1 - p_2$. Using the completeness relations for the Dirac and Rarita--Schwinger (Dirac) spinors with the help of the expression
\begin{equation}
    \begin{aligned}
        \sum_s u\inp{p} \bar{u}\inp{p} = & \slashed{p} + m \\
        \sum_s u_{\alpha}\inp{p} \bar{u}_{\beta}\inp{p} = & -\inp{ \slashed{p} + m} \inp{g_{\alpha\beta} - \frac{1}{3}\gamma_{\alpha}\gamma_{\beta} + \frac{2}{3} \frac{p_{\alpha} p_{\beta}}{m^2} + \frac{p_{\alpha} \gamma_{\beta} - \gamma_{\alpha} p_{\beta}}{3m}}
    \end{aligned}
    \label{371}
\end{equation}
one can, in principle, calculate the correlation functions from the hadronic part. However, at this point problems appear, which is generic for spin $3/2$-spin $1/2$ transitions. The problem is related to the fact that the interpolating current for spin $3/2$ baryons interacting not only with spin $3/2$ but also with spin $1/2$ negative parity baryons too.
Obviously, these pollutions should be eliminated. There are two different ways to overcome this problem. In the first method, one can introduce the projection operator $P_{3/2}$ which separates the contribution of spin $3/2$ states. The main idea of the second way for discarding the contributions of $1/2$ states is as follows. The matrix element for the interpolating current between vacuum and the particle with negative parity spin $1/2$ states can be written in general form as 
\begin{equation}
    \bra{0} \eta_{\mu} \ket{B\inp{p}} = \inp{A\gamma_{\mu} + B p_{\mu} } u\inp{p}~.
    \label{372}
\end{equation}
Multiplying both sides with $\gamma^{\mu}$ and using $\gamma_{\mu} \eta^{\mu} = 0$, we get $ 4 A + B m = 0$. From this expression, we obtain $\displaystyle A = -\frac{m}{4} B$; hence,
\begin{equation}
    \bra{0} \eta_{\mu} \ket{B\inp{p}} = B \inp{\gamma_{\mu} - \frac{m}{4} p_{\mu}} u\inp{p}~.
    \label{373}
\end{equation}

As a result, we observe that the structures proportional to $\gamma_{\mu}$ or $p_{\mu}$ contain the contributions of spin $1/2$ states only, and therefore these structures should be discarded. From Eq. \eqref{371}, we infer that $g_{\alpha\beta}$ structure contains the contribution of the spin $3/2$ states solely.

The second problem is that not all Lorentz structures are independent. This problem can be solved by using the specific order of the Dirac matrices. We choose the structure $\gamma_{\mu} \slashed{\varepsilon} \slashed{q} \slashed{p} \gamma_5$. 
Taking into account these remarks and using Eqs.~\eqref{370} and \eqref{371}, for the hadronic part of the correlation function, we obtain
\begin{equation}
  \begin{split}
    \Pi_{\mu}^{V} &= \frac{\lambda_1\lambda_2}{[m_1^2 - \inp{p+q}^2] [m_2^2 - p^2 ]} \bigg[ 
        -g_1 \inp{m_1+m_2}\slashed{\varepsilon}\slashed{p}\gamma_5 q_{\mu}
        +g_2 \, \slashed{q}\slashed{p} \gamma_5 \inp{p\varepsilon}q_{\mu}
        +g_3 \, m_V^2 \slashed{q}\slashed{p}\gamma_5\varepsilon_{\mu}
        + \parbox{4.5em}{other\\ structures} \bigg].
      \end{split}
\end{equation}

The QCD side of the correlation function is calculated in a standard manner by taking into account the vector meson DAs. The expressions of the DAs for the vector meson can be found in \cite{Ball:1998sk,Ball:1998ff,Ball:2004ye}. The values of the parameters in the expressions of the DAs for the light vector mesons at renormalization point $\mu = 1~\rm{GeV}$ are presented in Table~\ref{tab:7}. Our results for the the strong coupling constants of vector mesons with doubly heavy baryons after analyzing of the corresponding sum rules are presented in Table~\ref{tab:8}. The obtained results are crucial in analysis of the nonleptonic decays and can be used in the further studies of strong decays in the experiments carried out at LHC.

\begin{table*}[t]
  \centering
  \renewcommand{\arraystretch}{1.4}
  \setlength{\tabcolsep}{3.2pt}
  \begin{tabular}{lccccccccccc}
    \toprule
         Parameter & $\rho$ & $K^{*}$ & Parameter & $\rho$ & $K^{*}$ & Parameter & $\rho$ & $K^{*}$ & Parameter & $\rho$ & $K^{*}$ \\
    \midrule
        $f_V [GeV]$     & $0.216$ & $0.220$ & $a_1^{\parallel}$ & $0$    & $0.03$ & $\zeta_3^{\parallel}$           & $0.030$ & $0.023$  & $\zeta_4^{\parallel}$     & $0.07$  & $0.02$  \\
        $f_V^T [GeV]$   & $0.165$ & $0.185$ & $a_1^{\perp}$     & $0$    & $0.04$ & $\Tilde{\omega}_3^{\parallel}$  & $-0.09$ & $-0.07$  & $\zeta_4^{\perp}$         & $-0.08$ & $-0.05$ \\
        $m_V [GeV]$     & $0.770$ & $0.892$ & $a_2^{\parallel}$ & $0.15$ & $0.11$ & $\omega_3^{\parallel}$          & $0.15$  & $0.10$   & $\Tilde{\zeta}_4^{\perp}$ & $-0.08$ & $-0.05$ \\
                        &         &         & $a_2^{\perp}$     & $0.14$ & $0.10$ & $\omega_3^{\perp}$              & $0.55$  & $0.30$   & $\Tilde{\omega}_4^{\perp}$& $-0.03$ & $-0.02$ \\
                        &         &         &                   &        &        & $\kappa_3^{\parallel}$          & $0$     & $0$      & $\kappa_4^{\parallel}$    & $0$     & $-0.025$\\
                        &         &         &                   &        &        & $\kappa_3^{\perp}$              & $0$     & $0.003$  & $\kappa_4^{\perp}$        & $0$     & $0.013$ \\
                        &         &         &                   &        &        & $\lambda_3^{\parallel}$         & $0$     & $-0.008$ &                           &            &       \\
                        &         &         &                   &        &        & $\lambda_3^{\perp}$             & $0$     & $0.035$  &                           &             &      \\
                        &         &         &                   &        &        & $\Tilde{\lambda}_3^{\parallel}$ & $0$     & $-0.025$ &                           &   &                \\
    \bottomrule
  \end{tabular}
  \caption{The numerical values for the parameters in the DAs for the vector mesons $\rho$ and $K^{*}$. The renormalization scale is $\mu = 1~GeV$ \cite{Aliev:2020aon}.}
  \label{tab:7}
\end{table*}

\begin{table*}[t]
  \centering
  \renewcommand{\arraystretch}{1.4}
  \setlength{\tabcolsep}{3.2pt}
  \begin{tabular}{lccc|ccc}
    \toprule
        Transition & \multicolumn{3}{c|}{General current} & \multicolumn{3}{c}{Ioffe current}\\
    \midrule
        \tablexi{*}{cc}$\rightarrow$\tablexi{}{cc}$\rho^0$ & $ 1.13\pm0.25 $ & $ 0.11\pm0.03 $ & $ 7.81\pm1.83  $ & $ 0.99\pm0.22 $ & $ 0.1\pm0.02  $ & $ 6.92\pm1.62  $ \\
        \tablexi{*}{bb}$\rightarrow$\tablexi{}{bb}$\rho^0$ & $ 0.76\pm0.23 $ & $ 0.03\pm0    $ & $ 15.19\pm4.64 $ & $ 0.67\pm0.2  $ & $ 0.02\pm0    $ & $ 13.45\pm4.11 $ \\
        \tablexi{*}{bc}$\rightarrow$\tablexi{}{bc}$\rho^0$ & $ 1.06\pm0.2  $ & $ 0.05\pm0.01 $ & $ 14.44\pm2.84 $ & $ 0.94\pm0.18 $ & $ 0.05\pm0.01 $ & $ 12.79\pm2.51 $ \\
        \tablexi{*}{cc}$\rightarrow$\tablexi{}{cc}$\omega$ & $ 1.02\pm0.23 $ & $ 0.1\pm0.02  $ & $ 7.1\pm1.68   $ & $ 0.9\pm0.2   $ & $ 0.09\pm0.02 $ & $ 6.29\pm1.49  $ \\
        \tablexi{*}{bb}$\rightarrow$\tablexi{}{bb}$\omega$ & $ 0.69\pm0.21 $ & $ 0.03\pm0    $ & $ 13.82\pm4.25 $ & $ 0.61\pm0.19 $ & $ 0.02\pm0    $ & $ 12.24\pm3.77 $ \\
        \tablexi{*}{bc}$\rightarrow$\tablexi{}{bc}$\omega$ & $ 0.97\pm0.19 $ & $ 0.05\pm0.01 $ & $ 13.14\pm2.6  $ & $ 0.86\pm0.17 $ & $ 0.05\pm0.01 $ & $ 11.64\pm2.31 $ \\
        \tableom{*}{cc}$\rightarrow$\tableom{}{cc}$\phi$   & $ 1.5\pm0.32  $ & $ 0.5\pm0.14  $ & $ 9.79\pm2.5   $ & $ 1.32\pm0.28 $ & $ 0.45\pm0.13 $ & $ 8.64\pm2.22  $ \\
        \tableom{*}{bb}$\rightarrow$\tableom{}{bb}$\phi$   & $ 1.22\pm0.35 $ & $ 0.15\pm0.03 $ & $ 23.9\pm7.19  $ & $ 1.08\pm0.31 $ & $ 0.14\pm0.03 $ & $ 21.15\pm6.38 $ \\
        \tableom{*}{bc}$\rightarrow$\tableom{}{bc}$\phi$   & $ 1.47\pm0.29 $ & $ 0.25\pm0.04 $ & $ 19.26\pm4.13 $ & $ 1.3\pm0.26  $ & $ 0.22\pm0.03 $ & $ 17.03\pm3.66 $\\
    \bottomrule
  \end{tabular}
  \caption{The obtained values of the moduli of the coupling constants $g_1$, $g_2$, and $g_3$ for the given transitions accompanied by a light vector meson \cite{Aliev:2021hqq}.}
  \label{tab:8}
\end{table*}

\section{The vector dominance model and radiative decays of the doubly heavy baryons}
\label{sec:5}
In the previous section, we have determined the strong coupling constants of the light vector mesons with spin$\nicefrac{3}{2}$ and spin$\nicefrac{1}{2}$ doubly heavy baryons. Now using the values of the obtained coupling constants and the vector dominance model (VDM), we estimate the radiative decay widths of the $\Xi^*_{QQ}\rightarrow\Xi_{QQ}\gamma$ and $\Omega^*_{QQ}\rightarrow\Omega_{QQ}\gamma$.

The vector dominance model (VDM) implies that the form factors for the $B^*_{QQ}\rightarrow B_{QQ}\gamma$ transition can be obtained from $B^*_{QQ}\rightarrow B_{QQ}V$ transition form factors by converting the corresponding vector meson to photon. From the gauge invariance, the matrix element ${(\varepsilon^{\gamma *})}^\nu \bra{B^*_{QQ}(p)}j_\nu\ket{B_{QQ}(p+q)}$ is parameterized similar to ${(\varepsilon^{V*})}^\nu\bra{B^*_{QQ}(p)}j_\nu\ket{B_{QQ}(p+q)}$ in terms of three form factors
\begin{equation}
    \begin{aligned}
        (\varepsilon^{\gamma *})^\nu\bra{B*_{QQ}(p)}j_\nu\ket{B_{QQ}(p+q)} = e\bar{u}^\alpha(p,s)\Big[ & g_1^\gamma\left(\varepsilon^{\gamma*}_\alpha\slashed{q}-q_\alpha\slashed{\varepsilon}^{\gamma*}\right)
        +g_2^\gamma\left(Pq\varepsilon^{\gamma*}_\alpha-P\varepsilon^{\gamma*}q_\alpha\right)\\
        & + g_3^\gamma \left(q\varepsilon^{\gamma*}q_\alpha - q^2\varepsilon^{\gamma*}_\alpha\right)\Big]\gf u(p+q)~.
    \end{aligned}
    \label{374}
\end{equation}
Here $\varepsilon^{\gamma}$ is the polarization vector of the photon field. For the real photon case, obviously, the last term is equal to zero. To relate the form factors $g_i^\gamma$ with $g_i$ it is necessary to make the following replacement 
\begin{equation}
    \varepsilon_\mu \rightarrow \sum_{V=\rho,\omega,\phi} e\frac{f_V}{m_V}C_V\varepsilon_\mu^\gamma
    \label{375}
\end{equation}
and transfer from $q^2 = m_V$ to $q^2 = 0$.
Now let us check this ansatz. To calculate the matrix element $(\varepsilon^{V*})^\nu \bra{B^*_{QQ}(p)}j_\nu\ket{B_{QQ}(p+q)}$ we use the VDM ansatz and consider the contributions of intermediate vector mesons. In this case, we get
\begin{equation}
    \begin{aligned}
        \varepsilon^{\gamma*\nu}\bra{B^*_{QQ}(p)}j_\nu\ket{B_{QQ}(p+q)} = \sum_{V=\rho,\omega,\phi} \sum_\lambda e\frac{i}{q^2-m_V^2}(-i)\varepsilon^V_\lambda\varepsilon^{\gamma*}
        f_Vm_VC_V\varepsilon^{V*}_\lambda\bra{B^*_{QQ}(p)}j_\nu\ket{B_{QQ}(p+q)}~,
    \end{aligned} 
    \label{376}
\end{equation}
where $q^2 = 0$.
In this derivation, the form of the interaction Lagrangian of a vector meson with the photon is used as $\mathcal{L} = -ef_Vm_VC_VV^\mu A_\mu$  \cite{Zhao:2018mrg,Xing:2018lre}, where
\begin{equation}
    C_V = 
    \begin{cases}
        \frac{1}{\sqrt{2}} \text{ for } \rho \text{ meson}\\
        \frac{1}{3\sqrt{2}} \text{ for } \omega \text{ meson}\\
        -\frac{1}{3} \text{ for } \phi \text{ meson}~.
    \end{cases}
    \label{377}
\end{equation}
Once we take into account the quark contents of $\rho$, $\omega$ and $\phi$ mesons, the factor $C_V$ can be written in terms of the electric charges of corresponding quarks
\begin{equation}
    C_V = 
    \begin{cases}
        \frac{1}{\sqrt{2}}\left(e_u - e_d\right) \text{ for } \rho \text{ meson}\\
        \frac{1}{\sqrt{2}}\left(e_u + e_d\right) \text{ for } \omega \text{ meson}\\
        e_s \text{ for } \phi \text{ meson}~.
    \end{cases}
    \label{378}
\end{equation}
Then we use the definition of the matrix element $\varepsilon^{\gamma*\nu}_\lambda\bra{B^*_{QQ}(p)}j_\nu\ket{B_{QQ}(p+q)}$ and perform summation over the polarization of the massive vector mesons via
\begin{equation}
    \sum\varepsilon_\lambda^\mu\varepsilon_\lambda^{*\nu} = -g^{\mu\nu} + \frac{q^\mu q^\nu}{m_V^2}~.
    \label{379}
\end{equation}
Relying on the fact that in VDM, the vector meson's 4-momentum is equal to the photon's one, and using $\varepsilon^\gamma\cdot q = 0$, the last term in Eq. \eqref{379} can be neglected. As a result, we get
\begin{equation}
    \varepsilon^{\gamma*\nu}\bra{B^*_{QQ}(p)}j_\nu\ket{B_{QQ}(p+q)} = \sum_{V=\rho,\omega,\phi}e\frac{f_V}{m_V}C_V\varepsilon^{\gamma*}\bra{B^*_{QQ}(p)}j_\nu\ket{B_{QQ}(p+q)}~.
    \label{380}
\end{equation}
Comparing the coefficients of $\varepsilon^\gamma$ in both sides of Eqs.\eqref{370} and \eqref{380}, we get the following relation among the two sets of form factors
\begin{equation}
    g_i^\gamma = \sum_i e\frac{f_V}{m_V}g_i~.
    \label{381}
\end{equation}
Explicitly, these relations are written as
\begin{equation}
    g_i^\gamma = 
    \begin{cases}
        e_s\frac{f_\phi}{m_\phi}g_i^\phi \text{ for } \Omega^*_{QQ} \rightarrow \Omega_{QQ}\gamma\\
        \frac{1}{\sqrt{2}}e_u\left(\frac{f_\rho}{m_\rho}g_i^\rho \frac{f_\omega}{m_\omega}g_i^\omega\right) \text{ for } \Xi^*_{QQu} \rightarrow \Xi_{QQu}\gamma\\
        \frac{1}{\sqrt{2}}e_d\left(-\frac{f_\rho}{m_\rho}g_i^\rho \frac{f_\omega}{m_\omega}g_i^\omega\right) \text{ for } \Xi^*_{QQd} \rightarrow \Xi_{QQd}\gamma~.
    \end{cases}
    \label{382}
\end{equation}
Two remarks are in order. First, we assume that couplings do not change considerably when going from $q^2 = m_V^2$ to $q^2 = 0$. The second remark is related to the fact that the possibility of the heavy vector mesons contributing also. However, these contributions are neglected since in the heavy quark limit, they are proportional to $m_{J/\psi, \Upsilon}^{\nicefrac{-3}{2}}$.

From an experimental point of view, instead of the form factors $g_i^\gamma$ , the multipole form factors, such as magnetic dipole ($G_M$), electric quadrupole ($G_E$) and coulomb quadrupole ($G_C$) form factors are more convenient. The relations among $g_1^\gamma$, $g_2^\gamma$, $G_M$, $G_E$ and $G_C$ are given at Eq. \eqref{370} and at $q^2 = 0$, these relations are
\begin{equation}
    \begin{aligned}
        G_M =& \left(3m^*+m\right)\frac{m}{3m^*}g_1^\gamma + \left(m^*-m\right)\frac{m}{3}g_2^\gamma\\
        G_E =& \left(m^*-m\right)\frac{m}{3m^*}\left(g_1^\gamma + m^*g_2^\gamma\right)~.
    \end{aligned}
    \label{383}
\end{equation}
The decay width for $B_{QQ}^* \rightarrow B_{QQ}\gamma$ is given as
\begin{equation}
    \Gamma = \frac{3\alpha}{4}\frac{k_\gamma^3}{m^2}\left(3G_E^2 + G_M^2\right)~,
    \label{384}
\end{equation}
where $k_\gamma$ is the energy of the photon, $k_\gamma = \frac{m^{*2}-m^2}{2m^*}$.
The results for $G_E$ and $G_M$ are presented in Table \ref{tab:9}.
\begin{table*}[t]
  \centering
  \renewcommand{\arraystretch}{1.4}
  \setlength{\tabcolsep}{3.2pt}
  \begin{tabular}{lcc}
    \toprule
        Transition & $\left|G_E\right|$ & $\left|G_M\right|$ \\
    \midrule
        \tablexi{*++}{cc}$\rightarrow$\tablexi{++}{cc}$\gamma$ & $0.00\pm0.00$ & $1.78 \pm 0.4$\\
        \tablexi{*+}{cc}$\rightarrow$\tablexi{+}{cc}$\gamma$ & $0.00\pm0.00$ & $0.11 \pm 0.02$ \\
        \tablexi{*0}{bb}$\rightarrow$\tablexi{0}{bb}$\gamma$ & $0.00\pm0.00$ & $3.41 \pm 1.03$ \\
        \tablexi{*-}{bb}$\rightarrow$\tablexi{-}{bb}$\gamma$ & $0.00\pm0.00$ & $0.22 \pm 0.06$ \\
        \tableom{*+}{cc}$\rightarrow$\tableom{+}{cc}$\gamma$ & $0.00\pm0.00$ & $0.52 \pm 0.11$ \\
        \tablexi{*-}{bb}$\rightarrow$\tableom{-}{bb}$\gamma$ & $0.00\pm0.00$ & $1.18 \pm 0.34$ \\
    \bottomrule
  \end{tabular}
  \caption{The electric quadrupole and magnetic dipole form factors for the shown transitions \cite{Aliev:2021hqq}.}
  \label{tab:9}
\end{table*}

From the decay width expression, it follows that it is very sensitive to the mass difference $\Delta m = m^* - m$. A tiny change in mass difference leads to a significant variation for the decay width. For example, consider the decay width for $\Omega_{cc}^* \rightarrow \Omega_{cc}\gamma$ in our case (when $\Delta m = 0.86 MeV$) $\Gamma \sim 0.23 keV$. Our result is in good agreement with lattice QCD calculations~\cite{Brown:2014ena} but considerably different from the other approaches~\cite{LHCb:2019epo}.
One possible reason for this discrepancy is that VDM was not working quite well for doubly heavy baryon systems. In order to check how VDM works for doubly heavy baryon systems, it is necessary to calculate the radiative decays of them directly and compare the results with VDM predictions.

At the end of this section, we would like to add following remarks. As we already emphasized that the weak decays of double heavy baryons play a crucial role in deeply understanding the dynamics of doubly heavy baryons. First of all, the study of weak decay allows us to determine the CKM matrix elements more precisely and second, provides us with more useful information about QCD dynamics at a large distance. The semileptonic decays of doubly heavy baryons have very rich phenomenology and can be divided into two classes; 
\begin{itemize}
    \item Decays which take place at tree level of the charged current
    \item Decays which take place at loop level with flavor changing neutral current (FCNC).
\end{itemize}
The weak decays of doubly heavy baryons within the light-front quark model are intensively studied \cite{Wang:2017mqp,Zhao:2018mrg,Xing:2018lre}. Meanwhile, the weak decays via charged currents are studied within QCD sum rules~\cite{Shi:2019hbf} and light-cone sum rules \cite{Shi:2019fph,Hu:2019bqj}, respectively.
The weak decays via flavor-changing neutral current (FCNC) of doubly heavy baryons in the framework of light-cone sum rules have not been studied yet. These studies are critical for establishing the "right" picture of doubly heavy baryons. Because many nonleptonic decays are defined by the same transition form factors, these results are also quite timely for nonleptonic decays too. 
\section{Summary and conclusion \label{summary}}
The constituent quark model predicts the existence of baryons containing three quarks. However, although most of the baryons containing one heavy quark have already been observed, baryons with two or three heavy quarks not being discovered were a long-standing puzzle for the theory. The exciting news had been given by SELEX collaboration with the discovery of $\Xi_{cc}$ first, and confirmation was achieved by the LHCb collaboration after. While the experimental hunt has been going for discovering other baryons with doubly or triply heavy quarks, theoretical studies can also be enlightening for understanding the internal nature of these baryons and the determination of the fundamental properties as well as QCD parameters.

In this study, the properties of the doubly heavy baryons are investigated within the QCD sum rules, considering that these baryons are formed by a light quark attached to a doubly heavy diquark. We summarized the analyses conducted to determine the mass and strong coupling constants of the doubly heavy baryons. Moreover, our results obtained within QCD sum rules are compared with the other approaches.

Since most of the doubly heavy baryons predicted by the quark model have not been discovered yet, theoretical analysis can play a significant role in guiding the experiments. The experimental studies of the spectroscopic parameters and the weak decays of doubly heavy baryons constitute one of the essential directions in physical program of the colliders. Measurements of their decay characteristics can significantly extend our knowledge of the dynamics of these systems. From the theoretical point of view, their studies help develop new form of the heavy quark effective theory for doubly heavy baryons. Hopefully, the presented results on mass, strong coupling constants, and the electromagnetic decays of doubly heavy baryons will be used in further theoretical studies and will be tested at the LHCb experiment in the near future. 
%


\bibliographystyle{utcaps_mod}
\bibliography{../../all.bib}


\end{document}